\documentclass[twocolumn,aps,prl,showpacs]{revtex4-1}

\usepackage{graphicx,natbib,dcolumn,bm,amssymb,amsmath,latexsym,multirow,tabularx}

\begin{document}

\title{Nonlinear elasticity of semiflexible filament networks}

\author{Fanlong Meng and Eugene M. Terentjev}
\email{emt1000@cam.ac.uk}
\affiliation{Cavendish Laboratory, University of Cambridge, J. J. Thomson Avenue, Cambridge CB3 0HE, U.K.}

\begin{abstract}
\noindent We develop a continuum theory for equilibrium elasticity of a network of crosslinked semiflexible filaments, spanning the full range between flexible entropy-driven chains to stiff athermal rods. We choose the 3-chain constitutive model of network elasticity over several plausible candidates, and derive analytical expressions for the elastic energy at arbitrary strain, with the corresponding stress-strain relationship. The theory fits well to a wide range of experimental data on simple shear in different filament networks, quantitatively matching the differential shear modulus variation with stress, with only two adjustable parameters (which represent the filament stiffness and the pre-tension in the network, respectively). The general theory accurately describes the crossover between the positive and negative Poynting effect (normal stress on imposed shear) on increasing the stiffness of filaments forming the network.  We discuss the network stability (the point of marginal rigidity) and the phenomenon of tensegrity, showing that filament pre-tension on crosslinking into the network determines the magnitude of linear modulus $G_0$.  \\ 

\textit{Soft Matter}: Received 03 May 2016, Accepted 08 Jul 2016; doi: 10.1039/C6SM01029F
\end{abstract}

\maketitle

\section{Introduction}

\footnotetext{Cavendish Laboratory, University of Cambridge, Cambridge, CB3 0HE, UK.  E-mail: emt1000@cam.ac.uk}
\footnotetext{Soft Matter: Received 03 May 2016, Accepted 08 Jul 2016; doi:10.1039/C6SM01029F}

Networks of semiflexible filaments and fibers are common in biological systems, where dynamic structures of tunable strength, elasticity, and adjustable response are required \cite{Wickstead,Faury,Ozbek}.  From the microscopic scale of cytoskeleton to the macroscopic scale of fibrous tissue, such networks utilize stiff or semiflexible filaments, allowing for rich mechanical response \cite{Weitz2007}, dynamic remodelling \cite{Bursac,Fletcher},
and controlled structural failure \cite{Trepat}, while remaining an open structure allowing easy through access for solvents and solutes \cite{Lecuit}.
Inspired by these observations, filament networks are produced in industry,
and now stand as a promising area in manufacturing functional materials~\cite{CNT1,CNT2}.

The physics of  a single semiflexible chain or filament is well understood with the aid of worm-like chain model, in various implementations and approximations~\cite{Fixman,Marko1995,Ha1996,Bouchiat,Blundell2009}. The full theory is capable of accurately describing the force-extension relationship, as well as the magnitude of transverse fluctuations, for a range of filament stiffness spanning from very high (a rigid elastic rod) to very flexible (a classical polymer coil) and a range of end-to-end extensions from zero up to the full stretch where the entropic force has a characteristic divergence.
When such filaments form a macroscopic crosslinked network, the main scaling features of its nonlinear mechanical response are still determined by the single filament properties, as reviewed in~\cite{Macbook,Broedersz2014}.
There have been many key contributions to the elastic theory of stiff and semiflexible networks, including ones highlighting the issues of strain non-affinity~\cite{Head,Heussinger2007}.

In order to obtain the constitutive relationship of a semiflexible network, a widely used approach follows the following procedure (explained in detail in chapters 3,4 of ref.~\cite{DoiEdwards}): the shear stress on $ij$ plane is calculated by summing the contributions of the tension along the $i$ axis of all the chains crossing the $j$ plane. This approach has been very successful in deriving viscoelastic properties of polymer solutions and melts under shear flow~\cite{DoiEdwards,Morse1999}; Storm et al.~\cite{Storm2005} have applied the same stress-strain relation to a crosslinked network by assuming the affine elastic strain acting on each filament and the probability distribution of semiflexible filament length of Wilhelm and Frey~\cite{Frey1996}.  This model qualitatively describes how the network behaves when being deformed; however, it can only be used numerically (since no closed expressions are possible) and
it omits the pre-stress acting on the chains, as in this model the tension is assumed as zero when there is no deformation. Wilhelm and Frey also produced a numerical simulation of filament network elasticity on their own~\cite{Frey2003}, using the Mikado model of connectivity and athermal rigid rods as elastic elements. Although important issues of percolation rigidity threshold are exposed, this work cannot be used to describe most experimentally relevant (and most biological) filaments.
More recently, Palmer and Boyce~\cite{Boyce2008} proposed a closed analytical form of elastic free energy,
with the corresponding constitutive relation, using the same force-extension filament relationship as Storm et al.
They applied the `8-chain model' originally introduced in the context of ordinary rubber elasticity~\cite{Boyce2008} with an approximate expression for individual filament elasticity applied for each strand. This might be the best attempt in formulating the nonlinear elasticity of semiflexible network to date.
On the other hand, Unterberger et.al.~\cite{Unterberger2013} developed a `1-chain' model for a network, by assuming  filaments have a homogenous orientational distribution in the equilibrium system, and the average (imposed) stretch of the network $\lambda$ is a $p$-root average of deformations of individual filaments:  $\lambda=[\int (\lambda^*)^p \mathrm{d}A]^{1/p} / A^{1/p}$, where $\lambda^*(\theta, \phi)$ is the deformation of an individual filament, $p$ is the averaging parameter and $A$ the area of the unit sphere. This procedure allows the local stretch to fluctuate around its average value, and may partly account for the nonaffinity. This model could only be applied numerically to reproduce several key mechanical features (shear and normal stress).

Most good models discussed here make a successful fitting of shear stress data of different filament networks, which captures the generic stress-stiffening effect originating from the characteristic divergence of the force-extension curve of an individual filament. Usually, the authors employ the versions of celebrated Marco-Siggia interpolation formula~\cite{Marko1995}, which gives the correct response for filaments near full-extension (but much less so for more flexible filaments). In fact, the well-known MacKintosh scaling of differential shear modulus with stress, $K \propto \sigma^{3/2}$, in the stress-stiffening regime is entirely based on the mentioned divergence of an individual filament near full extension~\cite{Gardel2004}
(it holds even for an athermal network of undulating filaments, as long as the tensile force is proportional to the inverse-square of the compression, as is the case near full extension~\cite{Giessen08,Zagar2011}) . Therefore, a mere agreement (good fitting) of shear stress-strain curves is not a sufficient test of different theories. In particular, they must simultaneously descibe the effect of negative normal stress in the network of stiffer filaments, with a positive normal stress (also known as the Weissenberg effect) for networks made of more flexible chains~\cite{Janmey2007,Kang2009}. It turns out that none of the mentioned theories, including the 8-chain model of Palmer and Boyce, are able to produce the correct normal stress,
or address the network stability increasing with filament pre-tension\cite{Sharma2015}. In this paper we develop and discuss a continuum theory of semiflexible network (in closed analytical form) that addresses these issues, while retaining accurate fitting of a wide range of shear stress data.

\section{Network theory}
\begin{figure}
\centering
\includegraphics[width=0.99\columnwidth]{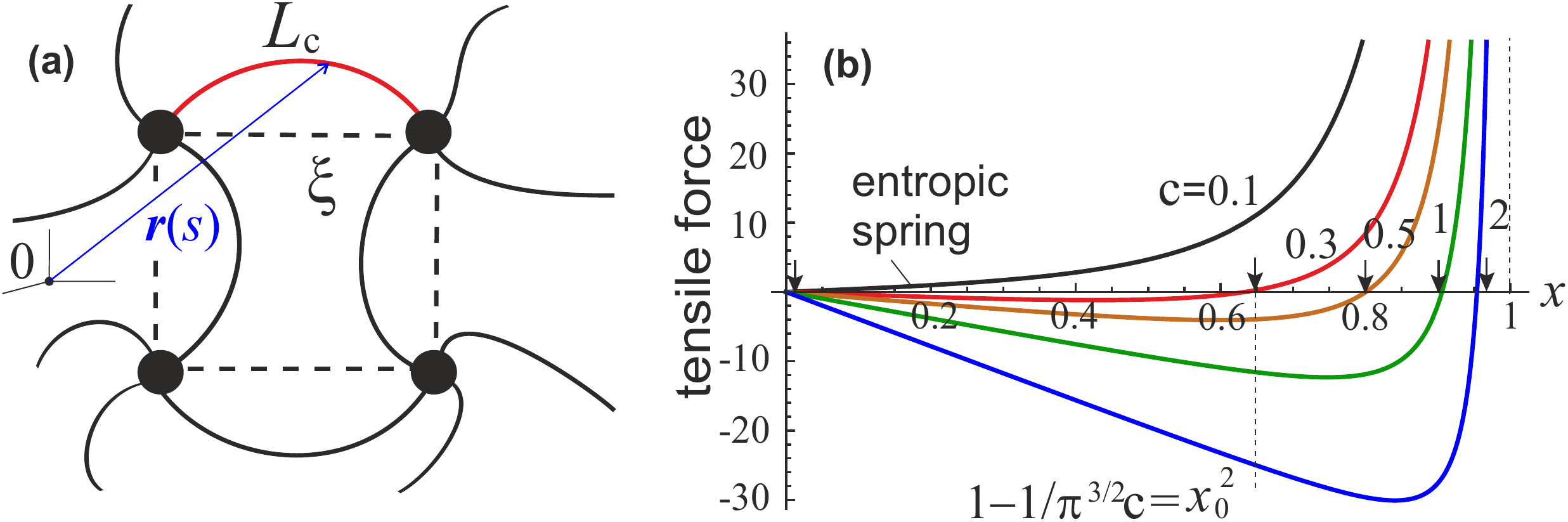}
\caption{(a) A sketch of semiflexible network mesh unit, with a bent filament connecting two crosslinks with its contour length  $L_\mathrm{c}$, and end-to-end length $\xi$ being the mesh size. \
(b) The relationship \cite{Blundell2009} between the tensile force, $f (x)/k_{\mathrm{B}}T$, and end-to-end factor, $x= \xi/L_\mathrm{c}$, plotted for semiflexible chains with different stiffness $c$, and marking the position $x_0$ for a chain to stay at the force-free state.}
\label{fig_1}
\end{figure}

Before introducing free energy of a semiflexible network, we need to discuss the properties of a single semiflexible filament.
A filament connecting two neighboring crosslinks in a network is sketched in Fig.~\ref{fig_1}(a), with
coordinates $\bm{r}(s)$, where $0\leq s \leq L_{c}$ is the arc-length coordinate along the chain. Different approaches to the chain (in)extensibility have been tested over the years~\cite{Vilgis1994}, from the strict constraint to the requirement that the length of filament remains constant on average (while small local fluctuations are allowed)~\cite{Frey1996,Blundell2009}, to the models that explicitly include filament stretching~\cite{Morse1998,Storm2005}. It turns out that in the regime of high extension, when there are no foldbacks (hairpins) on the chain, all length-constraining models give the same divergence of the force-extension, $f \propto 1/(1-x)^2$, with the end-to-end ratio $x=\xi / L_\mathrm{c}$, where $\xi$ is the end-to-end length of the filament, Fig. \ref{fig_1}(a). This was used by the famous interpolation formula of Marko and Siggia~\cite{Marko1995}. More recently, a complete theory of filament entropic elasticity has been developed, which spans the full range of extensions and the full range of bending modulus~\cite{Blundell2009}.

In the worm-like chain model~\cite{Kratky2010,Fixman},
the bending energy can be expressed by $\frac{1}{2} \kappa\int_{0}^{L_{c}} ds |\partial^{2} \bm{r}(s)/ \partial s^{2}|^2$, where $\kappa$ is the bending rigidity.
The key physical quantity for describing the stiffness of a polymer chain is the persistence length $l_p = \kappa/ k_\mathrm{B}T$.
When the contour length of the filament, $L_\mathrm{c}$, is comparable with its persistence length $l_{p}$, the chain is  regarded as ``semiflexible". Combining the effects of enthalpy arising from bending and entropy of conformation fluctuations, the closed form of the single chain free energy~\cite{Blundell2009} can be expressed as a function of its end-to-end factor, $x=\xi / L_\mathrm{c}$:
\begin{equation}\label{Achain}
F_\mathrm{chain} =  k_\mathrm{B} T \, \pi^2 c (1 - x^{2} ) + \frac{ k_\mathrm{B} T}{\pi c (1 - x^{2})} ,
\end{equation}
where  $c=\kappa / 2 k_\mathrm{B}TL_\mathrm{c}$ is a dimensionless stiffness parameter reflecting the competition between bending and thermal energy; in our notation  $c=l_p/2L_\mathrm{c}$. In ref.\cite{Blundell2009} one can find the comparison with several notable models of semiflexible filament (Marko-Siggia and Ha-Thirumalai) and where they deviate from the accurate expression (\ref{Achain}).

As shown in Fig~\ref{fig_1}(b),
if the value of $c$ is smaller than a critical value $c^{*}=\pi^{-3/2} \approx 0.18$,
the minimum of the free energy (or the force-free natural length) will be at $x=0$; such a chain can be regarded as flexible.
When $c \ll c^*$, one recovers the Gaussian entropic spring form: $F_\mathrm{chain} \approx  (2k_\mathrm{B} T /\pi l_p L_\mathrm{c}) \, \xi^2$. The Marko-Siggia (in fact, Fixman-Kovac) limit commonly used for quick fitting of force-extension curves is reached for flexible chains at high extension  $c \ll c^*,x \rightarrow 1$:  $F_\mathrm{chain} \approx  (k_\mathrm{B} T /\pi l_p L_\mathrm{c}) /(1-x)$.
When $c\gg c^{*}$, the chain can be regarded as a stiff rod with a natural length $x_0=\sqrt{1-c^{*}/c}$, and its energy is  dominated by elastic bending:   $F_\mathrm{chain} \approx (\pi^2 \kappa / L_\mathrm{c})[1-x]$. Under compression $x<x_0$ such a filament undergoes Euler buckling instability.  Note that on forming a crosslinked network, $x$ does not need to be equal to $x_0$ for each strand in equilibrium, meaning there can be pre-tension in the network.

We now construct the continuum elastic free energy of a network of such filaments using the methodology that was successfully developed in rubber elasticity~\cite{Boyce2000, Treloar}.
 If the sample shape is changed with a deformation tensor $\mathbf{E}$,
then the corresponding {Cauchy-Green} tensor~\cite{Bower} is $\mathbf{C}=\mathbf{E}\mathbf{E}^{\mathrm{T}}$,
with eigenvalues: $\lambda_{1}^{2}$, $\lambda_{2}^{2}$ and $\lambda_{3}^{2}$.
The values $\lambda_{1}$, $\lambda_{2}$, $\lambda_{3}$ can be interpreted as stretching ratios along the
principal directions of deformation. A class of simple, yet powerful theories of rubber elasticity is formulated in terms of invariants of the C-G tensor, $I_{1}=\lambda_{1}^{2}+\lambda_{2}^{2}+\lambda_{3}^{2}$,
$I_{2}=\lambda_{1}^{2}\lambda_{2}^{2}+\lambda_{2}^{2}\lambda_{3}^{2}+\lambda_{3}^{2}\lambda_{1}^{2}$
and $I_{3}=\lambda_{1}^{2}\lambda_{2}^{2}\lambda_{3}^{2}$.
Famous examples of this class are neo-Hookean (Gaussian), Mooney-Rivlin,  and Gent models (see \cite{Gent} for review).
Note that polymer networks are frequently treated as incompressible, so $I_{3}=1$.

\begin{figure} 
\centering
\includegraphics[width=0.99\columnwidth]{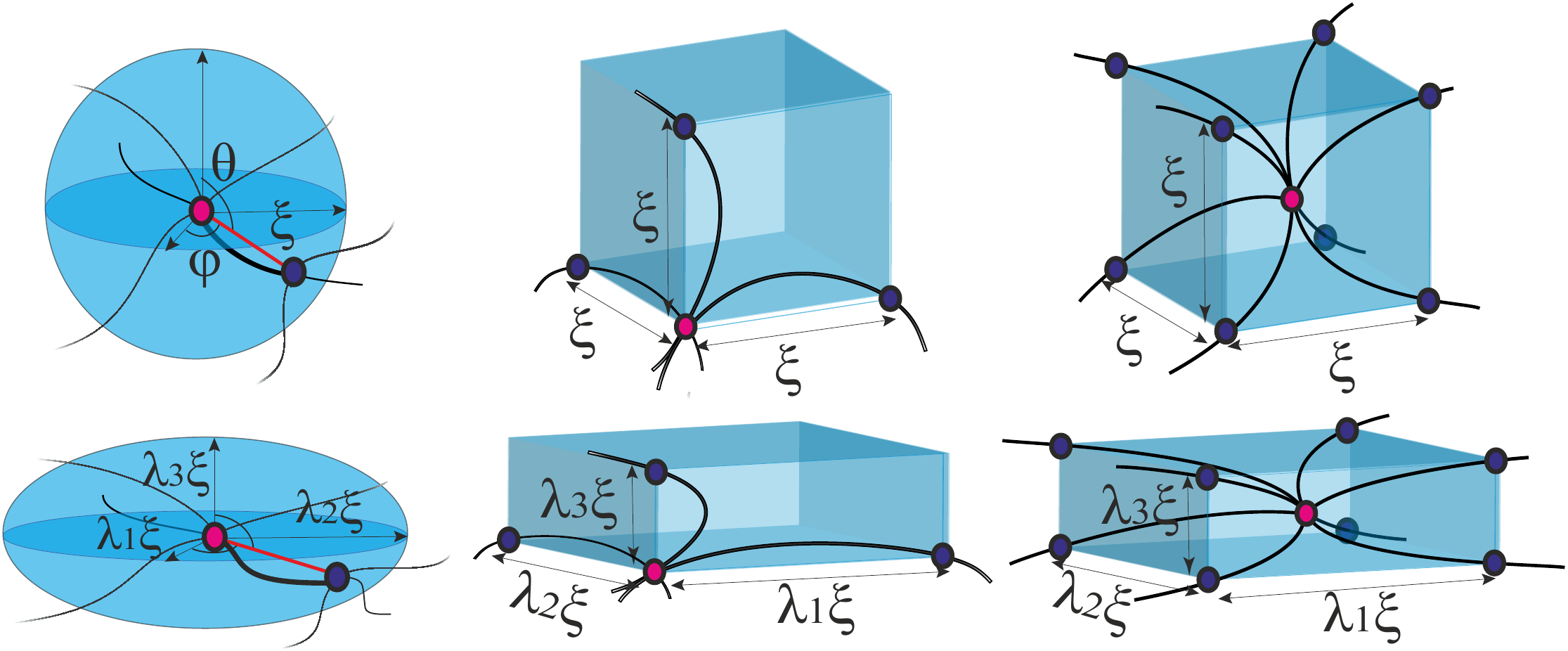}
\caption{Cells in a semiflexible network before and after deformations for the homogeneous sphere in full network model, primitive cubic lattice of 3-chain model, and  body-centred cubic lattice of 8-chain model.}
\label{models}
\end{figure}

For a polymer network, the chains are distributed and crosslinked randomly throughout the material,
which is the main difficulty in obtaining an analytical expression of the total free energy of the network (unless the individual chain is Gaussian). To incorporate a more complicated chain, such as Eq. (\ref{Achain}), into a constitutive framework, it is necessary to have a model that relates the chain deformation to the applied affine strain. This can be accomplished by representing an element of volume in the average network as a ``mesh cell'', which is then symmetrically multiplied to fill the volume.
Here we need to distinguish ``affinity" and ``non-affinity" in a filament network constructed by the unit-cell method.
A ``mesh cell" is deformed differently from the material, as these cells are constructed in orientation aligned with the principal stretching direction of the material,
which is referred to as ``non-affinity" by Palmer and Boyce~\cite{Boyce2008}. All mesh cells deform in the same way, with repeated deformed structure.
However, the real non-affinity can arise from the different responses among different ``mesh cells", due to local force relaxation.
In this work we refer to the concepts of ``affinity" or ``non-affinity" in the local context of individual filament junctions and mesh cells, rather than the global one between the orientation of a mesh cell and the macroscopic deformation of the material.
In assuming a symmetry of repeated mesh cells,
we automatically discard the effects of the local non-affine deformations~\cite{Zaccone,Liu}, i.e. take all mesh cells deforming uniformly, which we know must be the case at least for stiff filaments \cite{Head,Heussinger2007}.
Since there is no quantitative way of assessing the degree of the error introduced by the affine approximation (in our interpretation), we will have to look at the fits to the experimental data for validation. Within a mesh cell, chains or crosslinks have several possible arrangements, reflecting what one assumes about the topology of the network mesh, see Fig.~\ref{models}. Most acceptable structures include the  homogeneous sphere (HS) in 1-chain network model, the primitive cubic (PC) in 3-chain model, tetrahedral (TH) in 4-chain model, and body-centered cubic (BCC) in 8-chain model (see \cite{Boyce2008} and a review \cite{Boyce2000} for detail).

In 1-chain model~\cite{Boyce2000,Blundell2009,Unterberger2013},
one end of a polymer chain is fixed at the center of a sphere,
while the other end is on the sphere surface at an arbitrary orientation $(\theta, \varphi)$, distributed isotropically.
When deformed by $\mathbf{E}$, with stretching ratios along principal directions $\lambda_{1}$, $\lambda_{2}$, and $\lambda_3$, the lengths of the three semi-axes will change from $\xi$, to $\xi\lambda_{1}$, $\xi\lambda_{2}$ and $\xi\lambda_{3}$, respectively. The elastic energy density  of the network can be expressed as an orientational average of a deformed filament:
\begin{eqnarray}\label{fullnet}
F_\mathrm{1c} = n \int \frac{\sin \theta d\theta \, d\varphi}{4\pi} F_\mathrm{chain} [\tilde{\lambda} (\theta,\varphi ) \, \xi],
\end{eqnarray}
where $n$ is the density of crosslinked chains and $\tilde{\lambda}  (\theta,\varphi ) = \sqrt{\sin^{2}\theta(\cos^{2}\varphi\lambda_{1}^{2}+\sin^{2}\varphi\lambda_{2}^{2})+\cos^{2}\theta\lambda_{3}^{2}}$. If one takes the `entropic spring' limit of the Gaussian chain, the average $F_\mathrm{1c}$ reduces to the classical neo-Hookean rubber-elastic expression $n k_\mathrm{B}T (2\xi^2/3\pi l_p L_\mathrm{c})[\lambda_{1}^{2}+\lambda_{2}^{2}+\lambda_{3}^{2}]$.

However, in the general case the free energy in 1-chain model cannot be expressed in analytical form as a function of strain invariants, which renders it less convenient. In the following, we compare the 3- and 8-chain models, and decide on the 3-chain model preference, in particular due to the failure of 8-chain model in reproducing the normal stress.

In the 3-chain model, a primitive cubic is constructed with lattice points representing the crosslinking sites, and the edges are aligned along the principle directions of deformation tensor $\mathbf{E}$.
Three chains are linked with their end-to-end vectors along the edges and the equilibrium mesh size $\xi$.
On deformation, the lengths of three perpendicular edges at one lattice point become $\lambda_{1}\xi$, $\lambda_{2}\xi$ and $\lambda_{3}\xi$, respectively.
Then the free energy density of a semiflexible network can be expressed as
\begin{equation}
 F_{\mathrm{3c}} (\{ \lambda_{i=1,2,3} \} ) = \frac{n}{3} \sum_{i=1,2,3} F_\mathrm{chain} (\lambda_{i}\xi ). \label{3chain1}
\end{equation}
Equation~(\ref{3chain1}) can be rearranged as a function of the strain invariants:
\begin{eqnarray}\label{3chain2}
 F_{\mathrm{3c}}= \frac{n k_\mathrm{B} T}{3}\left[\pi^2 c \left(3-x^2 I_{1}\right)+
 \frac{3-2I_{1} x^2+I_{2} x^4}{\pi c\left(1-I_{1} x^2+I_{2} x^4-I_{3}x^6\right)}\right]
\end{eqnarray}
where $c=l_p/2L_\mathrm{c}$ and $x=\xi / L_\mathrm{c}$ as used in Eq.~(\ref{Achain}).

The body-centered cubic cell of 8-chain model is constructed with
eight filaments connected from the center point to all eight lattice points \cite{Boyce2008}. The edges of the cell are $\xi$ as we define the mesh size, while the chains are shorter by factor $\sqrt{3}/2$. The important feature of the high-symmetry 8-chain model is that on deformation all chains change their distance by exactly the same amount,  from  $\sqrt{3}\xi/2$ to simply $\xi\sqrt{I_{1}}/2$, since the mesh cell is aligned along the principal axes of deformation tensor $\mathbf{E}$. The free energy density of the network in this case is given by the single-chain expression directly:
\begin{eqnarray}\label{8chain}
 F_{\mathrm{8c}}(I_{1})&=& n F_\mathrm{chain}(\sqrt{I_{1}/3} \xi).
\end{eqnarray}

When the elastic energy is expressed as a function of strain invariants,
the stress tensor of an incompressible material (with $I_{3}=1$) can be obtained as~\cite{Gent,Bower}:
\begin{eqnarray}
 \sigma_{ij} &=& 2\left[\left(\frac{\partial F}{\partial I_{1}}+I_{1}\frac{\partial F}{\partial I_{2}}\right)C_{ij}-\left(I_{1}\frac{\partial F}{\partial I_{1}}+2I_{2}\frac{\partial F}{\partial I_{2}}\right)\frac{\delta_{ij}}{3} \right. \nonumber \\
&&  \qquad \left. -\frac{\partial F}{\partial I_{2}}C_{ik}C_{kj}\right] - P\delta_{ij},  \label{strain}
\end{eqnarray}
where $C_{ij}$ is the Cauchy strain, and $P$ the Lagrangian multiplier for incompressibility, the value of which determined by the boundary conditions.

\section{Simple shear deformation}
Let us consider the simple shear deformation such that in {Cartesian} coordinates a point  $(x,y,z)$ in an original material will change to $(x+\gamma z, y, z)$ after being sheared; $\gamma$ is the shear strain. The incompressibility is satisfied automatically, and the remaining strain invariants are: $I_{1} = I_{2} = 3+\gamma^{2}$.
The shear stress in 3-chain model can be obtained from the general constitutive relation (\ref{strain}) as:
\begin{eqnarray}\label{const3}
 \left[\sigma_{xz}\right]_{\mathrm{3c}} = \frac{2}{3}n k_\mathrm{B} T\gamma x^2 \left[
 \frac{\left(1-x^4\right)}{c \pi  \left[1-\left(2+\gamma^2\right) x^2+x^4\right]^2}-
  c \pi ^2\right],
\end{eqnarray}
while the corresponding expression in 8-chain model is:
\begin{eqnarray}\label{stress}
 \left[\sigma_{xz}\right]_{\mathrm{8c}}= \frac{2}{3}nk_\mathrm{B}T\gamma x^2\left[
 \frac{9}{c \pi  \left[3-\left(3+\gamma^2\right) x^2\right]^2}-
  c \pi ^2\right].
\end{eqnarray}
One must distinguish the nominal shear modulus $G(\gamma)=\sigma_{xz}(\gamma)/\gamma$,
from the differential shear modulus $K(\gamma)=\partial \sigma_{xz} /\partial \gamma$~\cite{Broedersz2014,Storm2005}, while the linear shear modulus of the network $G_{0}$ is equal to the differential modulus $K_{0}$ at $\gamma \rightarrow 0$.

\begin{figure} 
\centering
\includegraphics[width=0.99\columnwidth]{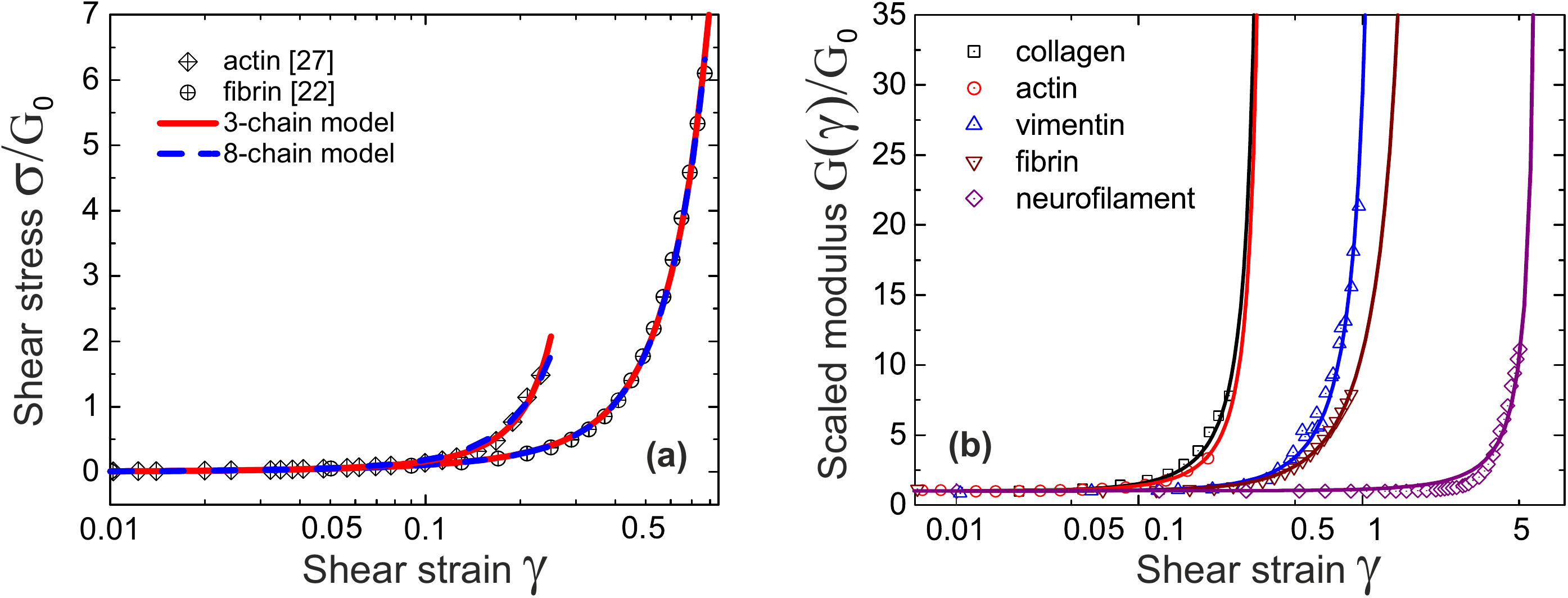}
\caption{(a) Fitting stress-strain experimental data of sheared actin~\cite{Gardel2004} and fibrin network~\cite{Storm2005} with 3-chain model (red curve) and 8-chain model (blue curve). \
(b) Fitting curves of experimental data~\cite{Storm2005} for sheared actin, collagen, vimentin, fibrin and neurofilament networks are optimally obtained for the scaled ratio $G(\gamma)/G_0$, which has a limit of 1 at $\gamma \rightarrow 0$; the fitting parameters for 3-chain model are given in Table~\ref{thenetwork}.}
\label{constitutiverelation}
\end{figure}

Take actin and fibrin network under simple shear deformation as an example.
Both 3-chain and 8-chain models fit the shear-experiment data from Refs.~\cite{Storm2005,Gardel2004} equally well, in fact -- perfectly, as shown in Fig.~\ref{constitutiverelation}(a).
The stiffness, $c$, and the initial end-to-end factor, $x=\xi/L_\mathrm{c}$, obtained by fitting with 3-chain model are a bit smaller than those in 8-chain model, but both in the semiflexible regime.
This is because the eight chains in a BCC are stretched equally (in principal axes) and share the deformation,
while the three chains in a PC are stretched differently and the most stretched one contributes the most to the nonlinear elastic energy.
We believe the intrinsic heterogeneity in the 3-chain model,
and the fact that the chain lying along the maximum principle stretch direction dominates the response of the whole cubic in semiflexible networks, is closer
to the realistic case of filament network. In addition, we shall see below that the 8-chain model cannot explain negative normal stress. Hence we apply the 3-chain model in the following. Figure~\ref{constitutiverelation}(b) shows fits to experimental data for a wide variety of semiflexible filaments.  In this figure we plot the shear modulus $G(\gamma)$ instead of stress, because the fitting convergence is much better when the initial data section is close to 1. Fitted parameters $(c,x)$ are listed in Table~\ref{thenetwork}, along with other parameters for each material that we list from the literature. 

\begin{table} 
\newcommand{\tabincell}[5]{\begin{tabular}{@{}#1@{}}#2\end{tabular}}
\begin{tabular}{|c|c||c|c||c|c|}
\hline
   & $G_0 ( \mathrm{Pa})$ & $c$ & $x $ & $l_{p} (\mu \mathrm{m})$ & $\xi (\mu  \mathrm{m})$\\
\hline
  collagen & 13.8  & 1.44 & 0.85 & 20.0 \cite{Collagen} & 5.9\\
\hline
  actin &  95.2 & 1.36 & 0.85 & 17.7 \cite{Gittes1993} & 5.5\\
\hline
  vimentin & 3.82 & 0.34 & 0.57 & 1.0 \cite{Mck2004} & 0.83\\
\hline
  fibrin & 18.9 & 0.25 & 0.40 & 0.50 \cite{Storm2005} & 0.40\\
\hline
  neurofilament & 2.83 & 0.14 & 0.15 & 0.45 \cite{Wagner2007}& 0.24\\
\hline
\end{tabular}
\caption{Fitting parameters $(c,x)$ for collagen, actin, vimentin, fibrin and neurofilament data obtained from \cite{Storm2005}. Also shown are the linear modulus $G_0$ extracted from the original data and used for scaling in Fig. \ref{constitutiverelation}, the literature values of $l_p$, and the calculated mesh size $\xi = l_p (x/2c)$.}
\label{thenetwork}
\end{table}

First of all, one may be surprised that the persistence length of collagen fibers is quoted as $20 \, \mu$m, when there is a large body of literature that would claim that collagen fibers are stiff athermal rods with persistence length of centimeters. This is all to do with the way a sample is prepared, and we use/quote the data~\cite{Collagen} where the collagen was apparently less aggregated than in a typical extra-cellular matrix. The same ambiguity will apply to actin networks as well, below. Another point to note about the fitted values in Table~\ref{thenetwork} is about neurofilament, which has $c < c^*$, that is, rather flexible chains. Taking the fitted value for $x=0.15 = \xi/L_c$, this leans that $L_c$ in this network was $\sim 1.6 \, \mu$m, i.e. about 3.6 times longer than $l_p$. By calculating $x_0=\sqrt{1-c^{*}/c}$ for different filaments listed in Table~\ref{thenetwork}, we see that the fitted $x$ is smaller than $x_0$,
indicating all of the filaments in Fig. \ref{constitutiverelation}(b) are pre-compressed, rather than pre-stretched in the equilibrium state of the network.

As Table~\ref{thenetwork} shows, no matter what the effective stiffness of the examined biofilaments,
the ratio of fitted parameters $c/x=l_{p}/2\xi$ remains close to 1. Later, when we discuss the network stability (Fig. 6), it will become clear that all the networks we examined here lie very near the stability boundary. It is not clear to us whether the fact that the mesh size is close to the filament persistence length  is an unintended result of different crosslinking density in experiments~\cite{Storm2005}, or is a relevant and universal biological feature.

\begin{figure} 
\centering
\includegraphics[width=0.99\columnwidth]{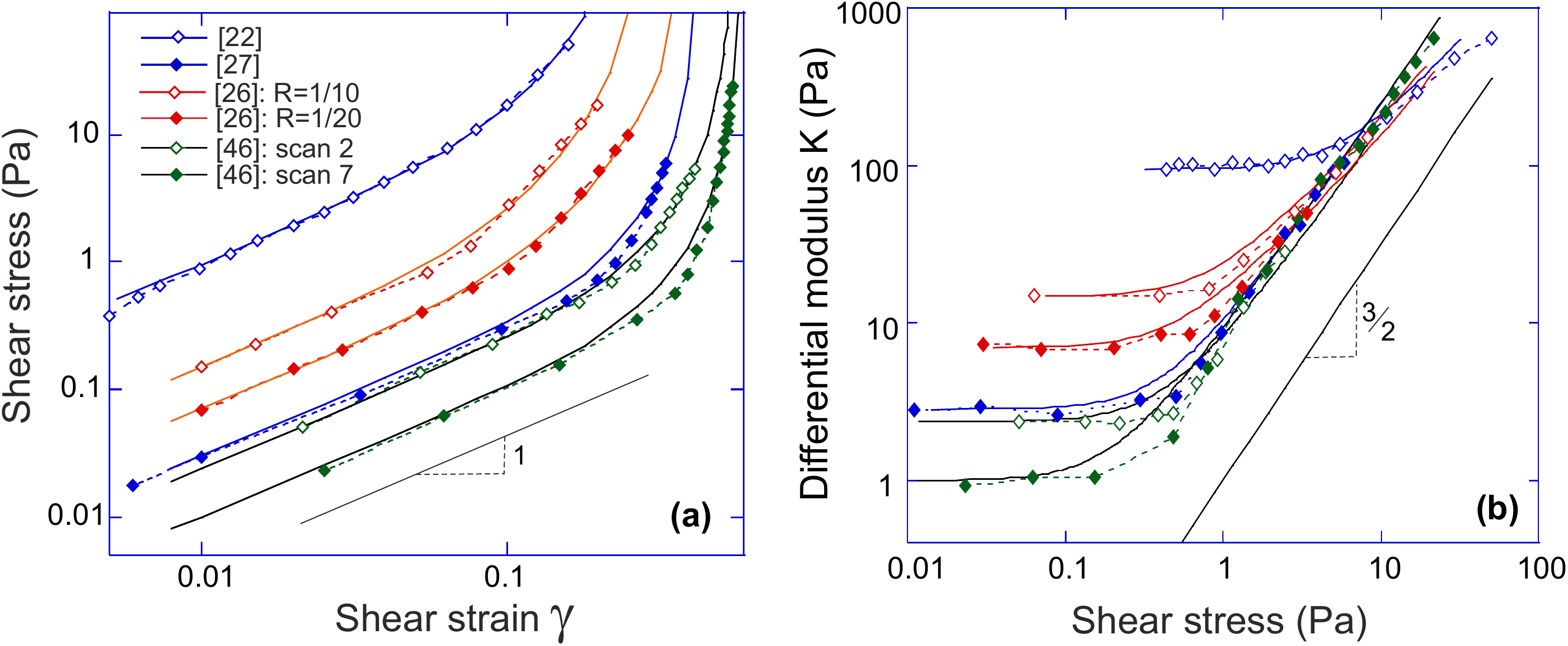}
\caption{(a) Fitting stress-strain experimental data of different actin networks under simple shear (the source references labelled in the plot).  The log-log plot nature highlights the linear-elasticity regime with the modulus $G_0$, before the stiffening sets in at higher shear. \
(b) The stress-stiffening of actin networks represented by the $K\sim \sigma^{3/2}$ scaling relation. All data sets are the same as labelled in the plot (a), and solid lines are theoretical curves plotted with Eq.~(\ref{const3}) using the same fitted parameters.}
\label{actin32}
\end{figure}

It is clear that networks of biological filaments have a great variety even within the same substance. Figure \ref{actin32} shows the published data for in-vitro crosslinked actin networks reported by different groups, all performing the simple-shear experiment (the data is digitized from a reference given in the plot). The stress-strain plotted in log-log format allows a clear identification of the linear regime $\sigma = G_0 \gamma$ (the modulus varying between 95\,Pa and 1\,Pa for different sets), and the subsequent stiffening at higher shear. All curves in Fig. \ref{actin32}(a) are fitted by the same Eq. (\ref{const3}) with the linear modulus $G_0$ and the two parameters $(c,x)$ taking values: $95$ Pa, $(1.43,\, 0.86)$; $14.8$ Pa, $(1.37,\, 0.85)$; $7.5$ Pa, $(0.94,\, 0.80)$; $3.0$ Pa, $(0.38,\, 0.80)$; $2.5$ Pa, $(0.38,\, 0.75)$ and $1.0$ Pa, $(0.27,\, 0.74)$ for the six sets from top to bottom. The difference between two data sets from \cite{Unterberger2013} is the density of heavy meromyosin (HMM) crosslinker (labelled on the plot); the difference between two data sets from  \cite{Schmoller10} is the degree of actin filament bundling: an initial network of F-actin filaments turned into a more sparse collection of bundles under repeated shear cycles \cite{Schmoller10}. As a result the network at `cycle 7' has lower $G_0$ but higher stiffening.

To find the onset of non-linearity in our theory, the differential shear modulus $K(\gamma)$ can be expanded in powers of shear strain $\gamma$:
\begin{eqnarray}
K(\gamma) & \approx &  \frac{2}{3} nk_\mathrm{B}T x^2 \left[\frac{1+x^2}{\pi c (1-x^2)^3}- \pi ^2 c \right]    \label{Ktaylor} \\
&& +4 nk_\mathrm{B}T x^4 \frac{1 + x^2}{c \pi (1 - x^2)^5} \, \gamma^2. \nonumber
\end{eqnarray}
The crossover point when these two terms are comparable with each other, in a more stiff network with $x \rightarrow 1$, can be approximated as
$\gamma_{t}\sim  (1 - x^2)/\sqrt{6}$. The stress at this point is $\sigma_{t} = G_{0} \gamma_{t} \sim 2 \sqrt{2}n k_\mathrm{B} T/[3\sqrt{3} c \pi (1 - x^2)^2]$.
This crossover stress $\sigma_{t}$ increases with the temperature as $(k_\mathrm{B}T)^2/\kappa$, closely matching the observations~\cite{Kouwer2013}.
At the end of range, when the shear strain approaches the point of divergence in stress $\sigma_{xz}$, ($\gamma \rightarrow 1/x -  x$ in Eq.~\ref{const3}), the stress and the differential modulus can be approximated as:
\begin{eqnarray}
  \frac{\sigma(\gamma)}{n k_\mathrm{B} T} &\simeq & \frac{2 \gamma x^2 (1 - x^4)}{
 3 \pi c \left[1 - (2 + \gamma^2) x^2 + x^4\right]^2}, \label{limit}  \\
  \frac{K(\gamma)}{n k_\mathrm{B} T} & \simeq & \frac{2 (1 - x^4) \left[x^2 - (2 - 3 \gamma^2) x^4 + x^6\right]}{
 3 \pi c \left[1 - (2 + \gamma^2) x^2 + x^4\right]^3}. \label{limit2}
\end{eqnarray}
Both expression diverge due to the same vanishing denominator, while maintaining the obvious scaling relation $K\sim \sigma^{3/2}$, which has been reported by different theories and experiments~\cite{Gardel2004a,Gardel2004,Lin2010,Lin2010a,Zagar2011,Kouwer2013}. We see this limit exposed clearly in Fig. \ref{actin32}(b) for very different actin networks. In this plot, the data points are the same as in Fig. \ref{actin32}(a), and we also plot the predicted curves of $K(\sigma)$ obtained by differentiating Eq. (\ref{const3}), using parameters $G_0, c$ and $x$ from the fitting in Fig. \ref{actin32}(a), for each data set.

One has to make a comment here, in the context of $K\sim \sigma^{3/2}$ scaling and different experiments. In many cases, such as in Unterberger et.al.~\cite{Unterberger2013}, the actin network was crosslinked by HMM and apparently retains some transient activity, producing the stress-softening and plasticity at higher stress (this is also the case in~\cite{Schmoller10} before the actin filaments formed bundles, or with F-actin crosslinked by filamin). Of course, our theory is not intended to deal with network plasicity (we assumed all crosslinkes permanent) and therefore we only retained the experimental data points in the early stress-stiffening regime to see the 3/2 scaling.

On the other hand, just by examining the actual data for the actin network of Storm et.al.~\cite{Storm2005}, one might conclude that it strongly and systematically deviates from the 3/2 scaling. In fact, the authors of~\cite{Storm2005} develop their own theory invoking various additional factors (e.g. filament extensibility) to account for this data. However, our basic theory, assuming permanent crosslinks, bulk incompressibility and inextensible filaments, evidently fits both $\sigma (\gamma)$ and $K(\sigma)$ data very well. The fact that the data (and the predicted curve) do not appear to follow the 3/2 scaling is be due to the fact that, for these values of $c$ and $x$, the final crossover to this characteristic scaling regime would occur at an even higher stress (at which point the network would probably not survive in practice).

\section{Normal stress}

\begin{figure} 
\centering
\includegraphics[width=0.99\columnwidth]{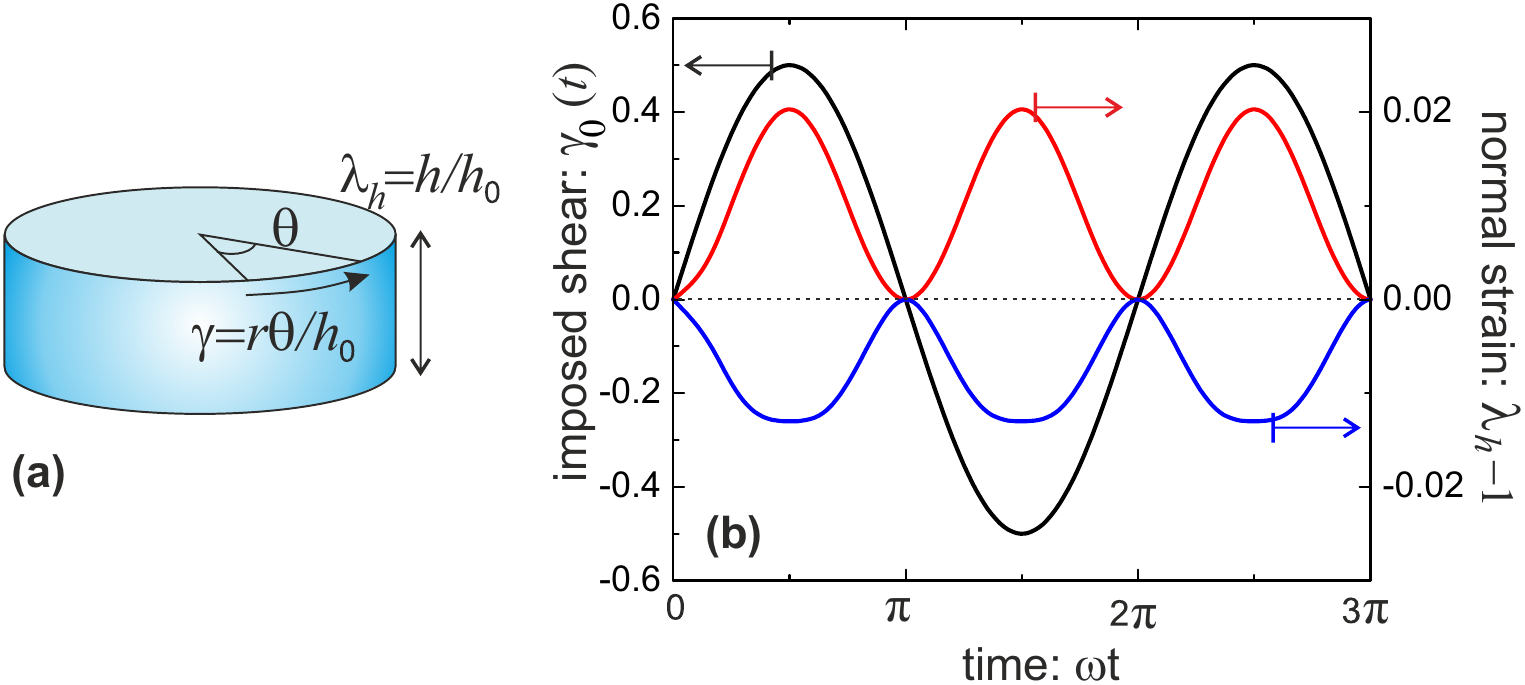}
\caption{(a) A cylinder-shaped sample under oscillating shear deformation; (b) the relationship between imposed oscillating shear strain $\gamma$ at the outer surface of the cylinder (black curve) and the normal strain $\lambda_h-1$ for a flexible network (red curve) with $c=0.1, x=0.1$, and a more stiff network (blue curve) with $c=0.3, x=0.6$.}
\label{poynting}
\end{figure}

Though simple shear deformation was helpful in this analysis, the important issue of normal stress remains controversial.
This is mainly because of the uncertain boundary conditions, see ref.\cite{Horgan2010} for detail. Since the experiments reporting normal stress measurements are most commonly conducted in rotating cylindrical geometry of a standard rheometer~ \cite{Janmey2007,Kang2009}, we will consider this geometry and realistic boundary conditions to describe the response of the material in its normal direction when a shear is applied, as shown in Fig.~\ref{poynting}(a).
Suppose the height and the radius of the undeformed cylinder are $h_{0}$ and $R_{0}$, respectively; on deformation
they may become $h=\lambda_{h}h_{0}$ and $R=\lambda_{R}R_{0}$. Incompressibility maintains  $\lambda_{h}\lambda_{R}^{2}=1$. In the $(r,\theta,z)$ coordinate system,  after rotating the top plate by the angle $\Theta$, the coordinates change as: $r \rightarrow r/\sqrt{\lambda_h}, \,
\theta \rightarrow \theta + \Theta z/h_{0}, \, z \rightarrow \lambda_h z$:
\begin{eqnarray}
\mathbf{E} =  \left(
                        \begin{array}{ccc}
                        1/ \sqrt{\lambda_{h}} & 0 & 0 \\
                         0 & 1/ \sqrt{\lambda_{h}} & \gamma(r)/\sqrt{\lambda_{h}} \\
                          0 & 0 & \lambda_h \\
                        \end{array}
                      \right),
\end{eqnarray}
where $\gamma(r)=r \Theta/h_{0}$ denotes the shear strain, which is a function of the radial position in this parallel-plate geometry.
The strain invariants become: $I_{1}=\lambda_h^2+[2+\gamma^2]/\lambda_h$
and $I_{2}=2 \lambda_h+[1+\gamma^2]/\lambda_h^{2}$. Given the shear strain at the outermost surface, $\gamma_{0}=\gamma(R_{0})$,
the total free energy then becomes a function of the stretching ratio $\lambda_h$ along the $z$ axis,
after integration over radius:
\begin{eqnarray}\label{energyofcylinder}
 F(\lambda_h; \gamma_{0}) = \int_{0}^{R_{0}}d r F_{\mathrm{3c}}(\lambda_h, r;\gamma_{0}).
\end{eqnarray}
The equilibrium $\lambda_h$ can be obtained by minimizing this free energy. When the cylinder radius becomes smaller (or larger) upon shear, with its height becoming correspondingly larger (or smaller) -- this phenomenon is called the positive (or negative) Poynting effect \cite{Mihai2011,Poynting}. This geometric effect with stress-free top plate corresponds to the positive (negative) normal stress required to maintain the fixed plate separation in a more common rheometry experiment \cite{Janmey2007,Kang2009}.
The 8-chain model fails in obtaining the correct normal stress in a sheared network:  due to its core assumption that eight chains in a cell are identically deformed (stretched) upon shear, the elongation ratio of the network along the stretching direction is always larger than 1, making the normal stress always positive.

Figure~\ref{poynting}(b) shows the results for equilibrium $\lambda_h$, presented as a response to an oscillating imposed shear $\gamma_{0}$, for two model materials with different filament stiffness and pre-tension. A flexible network ($c=0.1, x=0.1$) has a positive Poynting effect, or positive normal stress is required to counter the expansion along the height (in other setting, this is called the Weissenberg effect).
In a flexible network ($c<c^{*}$), when the mesh size is much smaller than the contour length of the subchains connecting the neighboring crosslinks,
\emph{i.e.},  $x_{0}\ll1$, the entropic energy plays a main role.
Though chains are stretched along the principal extension direction in the shear geometry illustrated in Fig.~\ref{poynting}(a),
the chains in other two principle directions are more likely being compressed, leading to the material contraction along the radial and circumferential directions.
However, if the mesh size is comparable with the contour length of the subchains, $0\ll x_{0}< 1$,
the material can behave in a negative Poynting effect manner:
the height of the cylindrical sample contracts, or negative normal stress is required to counter that and maintain the fixed height.
This is because the force acting on the chains directed along the principal extension in Fig.~\ref{poynting}(a) is
close to a divergence if $x_{0}\rightarrow 1$,
and it causes less energy when the material is compressed in the longitudinal direction,
rather than stretched.
Similar reason works also for a network of more stiff filaments (see red curve for $c=0.3 > c^*, x=0.6$).

We did not specifically calculate the normal stress here (this would be a cumbersome process involving the full tensor form of Eq. (\ref{strain}), first fixing the pressure $P$ from the condition of zero radial stress on the free outer surface). However, the magnitude of normal stress can be easily estimated from the linear relationship $\sigma_{h}\sim3G_{0}(\lambda_{h}-1)$, which uses the fact that the normal strain is quite small (i.e. the linear regime is justified) and the Young modulus is $3G_0$. Taking the fibrin values in Table~\ref{thenetwork} as an example, the normal stress is about $11.5$\,Pa under a shear strain of $\gamma_{0}=0.5$.
These are very close to the observations in experiments \cite{Janmey2007,Kang2009}. The magnitude of $\sigma_h \simeq 20$ Pa for the same shear also accurately matches the actin results obtained in \cite{Unterberger2013}.

\begin{figure} 
\centering
\includegraphics[width=0.59\columnwidth]{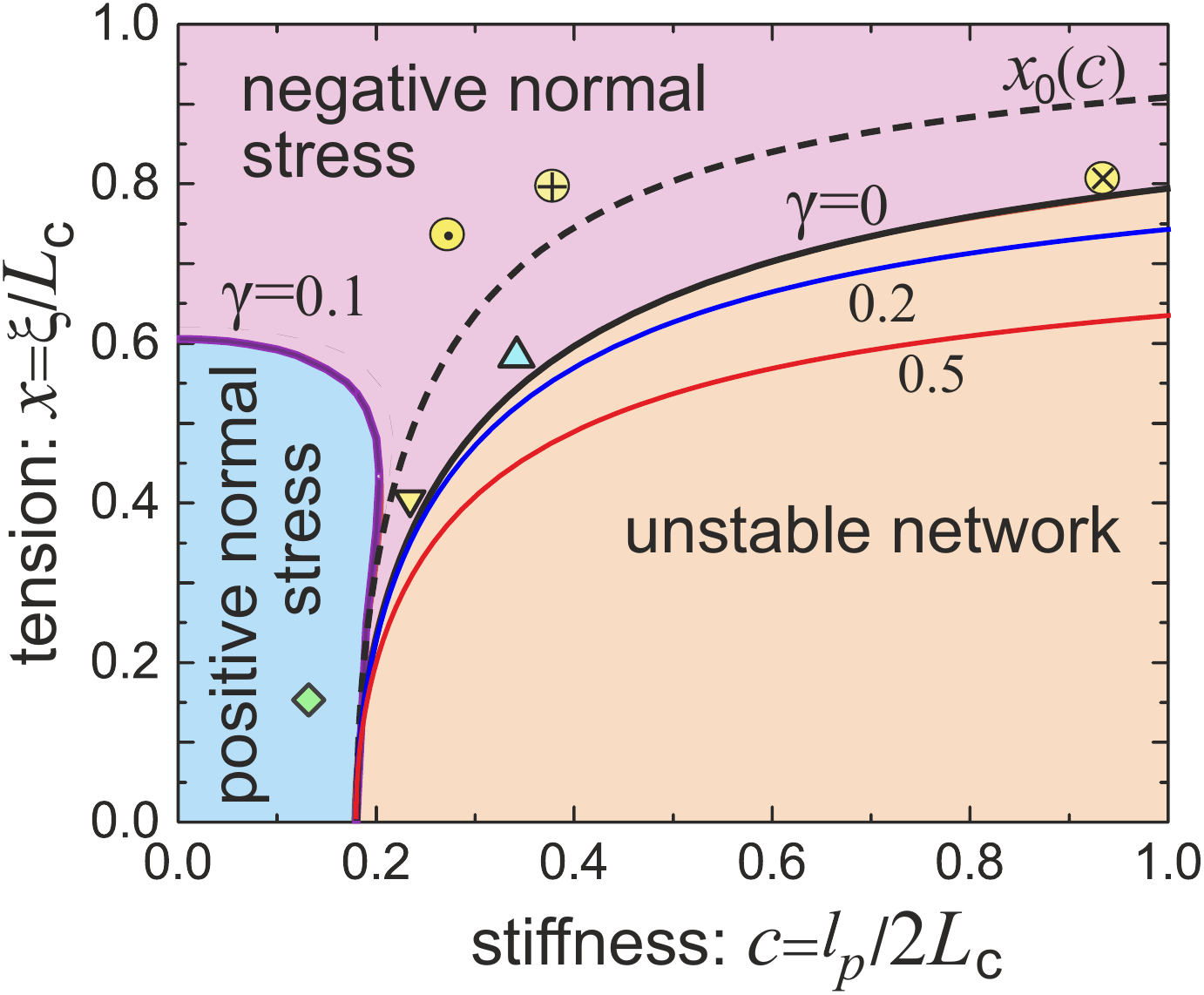}
\caption{The phase diagram of the network response at different stiffness, $c$, and filament pre-tension, $x$, showing the boundaries of positive/negative Poynting effect and the boundary of network stability (the dashed line of the neutral filament, $x_0$, was defined in Fig. \ref{fig_1}). Three `softer' filament networks from Table \ref{thenetwork} are shown in this map: $\triangle$ - vimentin, $\triangledown$ - fibrin, and $\diamond$ - neurofilament networks. Three actin networks from Fig. \ref{actin32} also fit on this map:  $\oplus$ - \cite{Gardel2004}, $\otimes$ - \cite{Unterberger2013}, and $\odot$ - \cite{Schmoller10}.  }
\label{map}
\end{figure}

Figure~\ref{map} gives the full `phase diagram' of the stiffness-tension parameter space ($c,x$) with phase boundaries separating positive/negative normal stress regions. In order to generate this diagram, for each parameter set ($c, x$) we have calculated the value of $\lambda_{h}$ by minimizing $F(\lambda_h; \gamma_{0})$, under the given shear. In this way the boundary separating the positive ($\lambda_{h}>1$, $\sigma_{h}>0$) and the negative ($\lambda_{h}<1$, $\sigma_{h}<0$) normal stress regions in Fig.~\ref{map} is calculated.
There is also a weak dependence of this boundary on the magnitude of the applied shear strain, which is due to the inherent non-linearity of stress-strain response; however, we are not showing this in the figure to avoid clutter.

Figure~\ref{map} shows that
a loose flexible network usually has a positive Poynting effect, while a network of more stiff filaments has a negative Poynting effect, especially when the filaments are crosslinked with increasing pre-tension.

\section{Network stability}

One can see from Figs.~\ref{constitutiverelation}(b) and \ref{actin32}(b) that the linear regime when $\sigma = G_0 \gamma$ persists for a different range of strain in different filaments; the onset of  hyperelastic regime is determined by how much pre-tension is in the filaments, i.e. how $x$ compares with $x_0(c)$  from Eq. (\ref{Achain}).
The linear shear modulus $G_{0}$ can be easily obtained from our theory: in 3-chain model it is given by the first term in Eq.~(\ref{Ktaylor}). The marginal rigidity condition $G_{0} \geq 0$ determines the strand pre-tension that is required for achieving a stable network. The stability criterion here is:
\begin{eqnarray}  \label{stabil}
   c \leq \frac{1}{\pi^{3/2}} \frac{ \sqrt{1+x^2}}{(1-x^2)^{3/2}} .
\end{eqnarray}
Note that the expression in the right hand side is always greater than $c^* = \pi^{-3/2}$ as defined in section 2.  Flexible chains with $c<c^*$ are therefore always stable in the rubbery network, that is, $G_0 >0$ always. On the other hand, stiff filaments have to be crosslinked with $\xi/L_\mathrm{c}$ exceeding the pre-tension threshold given by Eq.~(\ref{stabil}), which turns out to be slightly lower than $x_0(c)$ defined in Fig.~\ref{fig_1}(b). In other words, there have to be tensile forces acting on the crosslinked filaments in the network in order for it to be mechanically stable with a non-zero shear modulus. This notion is familiar from the ``tensegrity'' concept in biology and engineering~\cite{Tensegrity}. For stiff athermal filament network, the window of pre-tension between the linear modulus $G_0=0$ at $x \approx 1- 1/\pi  (2c)^{2/3}$, and $G_0\rightarrow \infty$ at $x \rightarrow 1$ (we assume inextensible chains) is very narrow.

The stability boundary of the network is plotted in Fig. \ref{map}. The condition for $G_0=0$ gives the equilibrium case labelled as $\gamma=0$ in the phase diagram. However, the full analysis shows that
the magnitude of the shear strain modifies  the stability condition, as represented by the coloured curves for $\gamma=0.2$ and $0.5$. This shift means that the region of mechanical stability of filament network expands on increasing deformation, which matches exactly what the recent paper by Sharma \textit{et al.} states~\cite{Sharma2015}.

To summarize, in this work we develop a continuum elastic theory of a network of semiflexible filaments,
by implementing the general free energy of one semiflexible chain into that of a disordered network. On reflection, we choose the 3-chain model as most closely matching the realistic system -- and achieve various quite stringent fits a number of different experimental data sets over the full range of nonlinear stress-strain range. We demonstrate that the general theory produces the  conditions for positive/negative Poynting effect (normal stress) in a network under imposed shear. The greatest weakness of this model is omitting the effects of local strain non-affinity, which might play an important role when crosslink density of the network is small and filaments stiff. The effect of tensegrity, or linking of filament pre-tension to the network stability and the magnitude of the shear modulus $G$ is an unexpected result of this theory. The stringent match of experimental data is reassuring, and we believe that the presented analytical continuum model can provide as an efficient and portable  tool for studying the mechanical properties of semiflexible networks. \\

\noindent \textbf{Acknowledgments:} \ 
This work has been funded by the TCM Critical Mass Grant from EPSRC (EP/J017639).
We are grateful for informative discussions with Masao Doi and Zhongcan Ouyang,
and thank Cornelius Storm for providing their raw experimental data \cite{Storm2005}.


\providecommand*{\mcitethebibliography}{\thebibliography}
\csname @ifundefined\endcsname{endmcitethebibliography}
{\let\endmcitethebibliography\endthebibliography}{}

\end{document}